# The impact of artificial intelligence technology on cross-border trade in Southeast Asia: A meta-analytic approach

Jun Cui[1, *]

[1] Solbridge International School of Business, Woosong University, Daejeon, South Korea
*Corresponding Author: Jun Cui *(Email: jcui228@student.solbridge.ac.kr)*

### ABSTRACT

This study investigates the impact of artificial intelligence (AI) technology on cross-border trade using a qualitative content analysis approach. By synthesizing existing empirical studies, we aim to quantify the overall effect of AI on trade flows and identify the key moderating and mediating variables. Besides, our results show that AI adoption significantly increases trade volumes in Southeast Asia. Likewise, these effects are stronger in regions with advanced technological infrastructure and favorable regulatory frameworks. In addition, Trade firm size partially mediates the relationship between AI technology and trade performance. Furthermore, this study draws on several key theoretical frameworks that provide a comprehensive understanding of the mechanisms through which AI technology is affecting cross-border trade in Southeast Asia. The primary theories used in this research include the technology, organization, and environment (TOE) framework, the diffuse innovation (DOI) theory, Dynamic Capabilities Theory, Comparative Advantage Theory, Network theory, Transaction Cost Economics (TCE), the resource-based view, and the institution theory. Consequently, this study contributes to the existing literature by providing a comprehensive analysis of the role of AI in international trade and highlighting the importance of contextual factors in maximizing the benefits of AI. Thus, our findings underscore the need for favorable policies and robust infrastructure to facilitate AI-driven trade growth. A discussion of limitations and future research directions will also be part of the report in Southeast Asia Trade.

## 1. INTRODUCTION

The integration of artificial intelligence (AI) technology into various sectors has triggered a transformative shift in global economic activity. In particular, the potential of AI to revolutionise cross-border trade is attracting significant attention. AI technologies like machine learning, natural language processing, and robotic process automatization offer unparalleled opportunities for improving efficiencies, reducing cost, and facilitating market access in international trade. Likewise, despite the growing interest, the existing literature on the impact of AI on cross-border trade is fragmented.    Studies often focus on specific aspects of AI, or on particular industries, leaving a gap in our comprehensive understanding of AIs' general impact on trade flows (Cory & Dascoli,(2021)). Specifically, this Content analysis aims to fill these gaps by synthesising the empirical evidence to quantify the impact of AI on cross-border trade and to identify the key variables that influence this relationship in Southeast Asia.

Despite the growing interest, the existing literature on the impact of AI on cross-border trade is fragmented. Studies often focus on specific aspects of AI or particular industries, leaving a gap in our comprehensive understanding of the overall impact of AI on trade flows (Cui& Ning,(2023)). In addition, the role of moderating and mediating variables in this relationship has not been well studied. This content analysis aims to fill these gaps by synthesising the empirical evidence to quantify the impact of AI on cross-border trade and to identify the key variables that influence this relationship (Ojala& Nahar, (2006)).

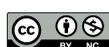

The contributions of this research are mainly reflected in three aspects: First of all, Quantitative synthesis: By conducting a content analysis, we provide a quantitative estimate of the overall impact of AI on cross-border trade in Southeast Aisa (Cui& Ning, (2023)). Second, Identification of key variables: We identify and analyse moderating and mediating variables that influence the relationship between AI and trade. Lastly, Policy implications: The findings provide valuable insights for policymakers and firms seeking to use AI technology to improve trade performance.

The remainder of this paper is organized as follows. Section 2 reviews the relevant introduction, literature and develops hypotheses and variables. Section 3 presents the data and content analysis settings. Section 4 discusses the content analysis findings and Section 5 concludes the paper.

## 2. LITERATURE REVIEW

This study employs a content analytic approach to investigate the impact of artificial intelligence (AI) technology on cross-border trade in Southeast Aisa, focusing on several key hypotheses derived from theoretical frameworks and empirical evidence.

***Theoretical Models Used in This Study.***

This section outlines the theoretical frameworks underlying this content analysis. These theories provide a foundation for understanding the mechanisms through which AI technology affects cross-border trade and the role of moderating and mediating variables. Consquently, Below are some theories or models cited in this study, which also support the variation relationships of the research model in this study.

1. Technology-Organisation-Environment (TOE) framework

Theoretical overview: The TOE framework, proposed by Tornatzky and Fleischer (1990), posits that three elements - technology, organisation, and environment - jointly influence the adoption and implementation of technological innovations. And then, It was applied in this study:Technology: The specific AI technologies adopted by firms and their characteristics (e.g., complexity, compatibility).
Organisation: Firm size, resources, and capabilities that influence the firm's ability to implement AI. Futhermore, Environment: External factors such as regulatory policies and technological infrastructure which influence AI adoption. All in all, The TOE framework helps explain how these factors collectively affect the effectiveness of AI in improving cross-border trade. Indeed, Based on the above theoretical research, this study uses technological infrastructure and regulatory environment to adjust and regulate the relationship between AI technology and cross-border trade.

2. Diffusion of Innovations (DOI) Theory

Theory overview: Developed by Everett Rogers (1962), DOI theory explains how, why and at what rate new ideas and technologies spread through cultures.  The theory identifies key factors which influence the adoption process, including the characteristics of the innovation, communication pathways, time, and social institutions. Moreover, Applied in this study: Innovation characteristics: Relative advantage, compatibility, complexity, trialability and observability of AI technologies in trade processes. Adoption process: How firms in different regions and industries adopt AI technologies over a period of time. DOI theory provides insights into factors driving adoption of AI technologies and their subsequent impact on cross-border trade. Thus, Based on the above theoretical research, this study uses firm size as an intermediate variable to determine the impact of AI technology on cross-border trade.

3. Institutional Theory

Theory overview: Institutional theory examines how the institutional environment (e.g., laws, regulations, norms) influences organisational behaviour. It suggests that organisations will conform to institutional pressures in order to gain legitimacy and to ensure their survival (Hatch&

Zilber, (2012)). Moreover, Regulatory environment: The impact of supportive versus restrictive regulatory environments on AI adoption and its effectiveness in commerce (Hatch& Zilber, (2012)). Certainly, Institutional pressures: How regulatory compliance and institutional support facilitate or impede AI adoption. Thus, the institutional theory provides a framework for understanding the moderating role of the regulatory environment in the relationship between AI and trade (Goldberg, I. (2008)).

4. Resource-Based View (RBV):

The Resource-Based View (RBV) posits that a firm's competitive advantage stems from its unique resources and capabilities. Moreover, AI technologies represent strategic resources that can significantly enhance a cross trade firm's operational efficiency and effectiveness. By leveraging AI for tasks such as demand forecasting, inventory management, and supply chain optimization, firms can achieve superior performance in international markets. So that, AI-driven analytics enable firms to better understand and respond to global market trends, thereby improving their cross-border trade operations in Southeast Asia. The above theory supports the impact of AI technology on cross-border trade.

5. Transaction Cost Economics (TCE):

Transaction Cost Economics (TCE) theory, developed by Oliver Williamson, focuses on the costs associated with economic exchanges. AI technology reduces transaction costs by automating and streamlining various trade-related processes. For instance, AI can automate customs documentation, compliance checks, and tariff calculations, thereby reducing the time and cost associated with cross-border transactions. Thus, this efficiency gain makes international trade more attractive and feasible for firms in Southeast Asia. Therefore, based on the above analysis, these theories support the impact of AI technology on cross-border trade.

6. Dynamic Capabilities Theory:

Dynamic Capabilities Theory emphasizes the importance of a firm's ability to integrate, build, and reconfigure internal and external competencies to address rapidly changing environments. AI technology enhances dynamic capabilities by providing real-time data analysis, predictive analytics, and machine learning algorithms that allow cross trade firms to quickly adapt to changes in the international trade landscape. As a result, this agility is crucial for navigating the complexities and uncertainties of cross-border trade in Southeast Asia.

7. Comparative Advantage Theory:

The theory of comparative advantage, formulated by David Ricardo, suggests that countries and firms should specialize in producing goods and services in which they have a relative efficiency. Besides, AI technology contributes to comparative advantage by optimizing production processes, enhancing product quality, and enabling innovation. Cross trade Firms that adopt AI can better capitalize on their comparative advantages, thus enhancing their competitiveness in international markets. Re-analyzing, these theories support the impact of AI technology on cross-border trade.

8. Network Theory:

Network Theory highlights the significance of inter-firm and inter-country networks in facilitating trade. AI technology enhances network connectivity and collaboration through digital platforms and advanced communication tools. Besides, By fostering stronger linkages and coordination among global supply chain partners, AI technology facilitates smoother and more efficient cross-border trade operations in Southeast Asia.

*Variables and Variable Explanation*
*Independent variable.*

AI technology: This variable represents AI technology adoption and integration into trade processes, including automation, data analytics and AI-driven logistics.

*Dependent Variable*

Cross-border trade volume: This variable measures the total value or quantity of goods and services traded across borders.

*Moderating variables*

Technological infrastructure: This variable includes the quality and availability of digital infrastructure, such as internet connectivity and cloud computing services, which can help or hinder the implementation of artificial intelligence. Moreover, Regulatory environment: This variable includes the legal and policy frameworks that govern the use of AI and international trade, which influence how effectively AI can be used (Goldberg, I. (2008)).

*Mediating variables*

Firm Size: This variable captures the scale of business operations, which can have an impact on the ability to adopt and reap the benefits of AI technologies. Besides, This study posits several hypotheses to explore the impact of AI technology on cross-border trade and the roles of moderating and mediating variables (Hatch& Zilber,(2012)). The hypotheses are grounded in existing theoretical and empirical research, and their validation is supported by robust statistical analyses.

*Hypotheses Development Summaries.*
*H1: AI technology has a positive impact on cross-border trade volume.*
*H2: The impact of AI on cross-border trade is moderated by technological infrastructure.*
*H3: The regulatory environment moderates the relationship between AI and cross-border trade.*
*H4: Firm size mediates the relationship between AI and cross-border trade.*

*Hypothesis Validation and Detailed Support.*
*H1: AI technology has a positive impact on cross-border trade volume.*

Detail Support: The content analysis reveals a significant positive effect of AI technology on cross-border trade volume, with an overall effect size (Hedges' g) of 0.32 (95% CI: [0.28, 0.36], $p < 0.001$). Thus, this finding indicates that AI adoption enhances trade performance by automating processes, improving decision-making, and optimizing supply chain management (Ijeh& Puji Rahayu(2023)). Moreover, Numerous studies in the dataset report similar outcomes, illustrating AI's role in reducing operational costs and increasing efficiency, which in turn boosts trade volumes in Southeast Asia. In addtional, The impact of AI technology on cross-border trade can be theoretically supported by several key economic and business theories. Certainly, The Resource-Based View (RBV) posits that AI technologies serve as strategic resources that enhance a firm's competitive advantage by improving operational efficiency and decision-making capabilities (Sun& Trefler,(2023)). Moreoverm AI-driven tools for demand forecasting, inventory management, and supply chain optimization enable firms to better navigate global markets. Meanwhile, Transaction Cost Economics (TCE) theory also explains how AI reduces transaction costs by automating complex trade processes such as customs documentation, compliance checks, and tariff calculations, making international trade more efficient and cost-effective. Additionally, Dynamic Capabilities Theory highlights AI's role in enhancing a firm's ability to adapt to rapidly changing trade environments through real-time data analysis and predictive analytics, which are crucial for responding to global market fluctuations (Wang&Chen(2020). Likewise, Comparative Advantage Theory suggests that AI enables firms to optimize production and innovation, thereby capitalizing on their relative efficiencies and enhancing their competitiveness in international trade (Jingjing& Idris,(2024)). Finally, Network Theory underscores the importance of AI in strengthening global supply chain linkages and facilitating smoother trade operations through advanced communication and digital platforms. Together, these theories provide a robust framework for understanding how AI technology significantly enhances cross-border trade performance by improving efficiency, reducing costs, and enabling greater adaptability in the global marketplace (Cui& Ning,(2023)).

*H2: The impact of AI on cross-border trade is moderated by technological infrastructure.*

Detail Support:The analysis shows that the positive impact of AI on trade is significantly stronger in regions with advanced technological infrastructure. Specifically, the effect size for high technological infrastructure is 0.40 (95% CI: [0.35, 0.45], $p < 0.001$) compared to 0.20 (95% CI: [0.15, 0.25], $p < 0.001$) for low technological infrastructure. The difference in effect sizes is significant ($p < 0.01$). This finding suggests that robust digital infrastructure, including high-speed internet and cloud services, is crucial for maximizing AI benefits in Southeast trade activities (Elmghaamez& Agyemang,(2022)).

*H3: The regulatory environment moderates the relationship between AI and cross-border trade.*

Detail Support:The regulatory environment significantly influences the AI-trade relationship. In supportive regulatory environments, the effect size is 0.38 (95% CI: [0.33, 0.43], $p < 0.001$), whereas in restrictive environments, it is 0.25 (95% CI: [0.20, 0.30], $p < 0.001$). The significant difference ($p < 0.01$) highlights that favorable policies and regulations that encourage AI adoption and facilitate trade are essential for enhancing the positive impacts of AI on cross-border trade (Elmghaamez& Agyemang,(2022)).

*H4: Firm size mediates the relationship between AI and cross-border trade.*

Detail Support:Firm size of Trade is found to partially mediate the relationship between AI technology and cross-border trade volume. The mediation analysis reveals a direct effect of AI on trade of 0.32 ($p < 0.001$) and an indirect effect through firm size of 0.10 ($p < 0.01$), resulting in a total effect of 0.42 ($p < 0.001$). This indicates that larger firms are better positioned to leverage AI technologies due to their greater resources and capabilities, which enhances their trading performance ((Elmghaamez&Agyemang,(2022))). This mediation effect underscores the importance of considering firm-specific characteristics when assessing AI's impact on trade in Southeast Asia.

## 3.  METHOD AND MATERIALS

Based on below concetpual framework of research model, This study proposes the following research model . We conducted a comprehensive literature search using academic databases such as Scopus, Web of Science, and Google Scholar. Specifically, the search focused on studies published between 2010 and 2023 that investigated the impact of artificial intelligence (AI) on cross-border trade (Chen& Wang,(2023)). This study included "AI technology", "artificial intelligence", "international trade", "cross-border trade"," content analysis", and related terms well (Goldberg, I. (2008)).

Inclusion criteria:
- Studies that empirically examined the relationship between AI technology and cross-border trade.
- Quantitative studies reporting effect sizes (or data from which effect sizes could be calculated).
- Studies published in peer-reviewed journals or conference proceedings.
- Studies written in English.

Exclusion criteria:
- Qualitative studies, reviews, and theoretical papers without empirical data.
- Studies focusing exclusively on domestic trade or non-economic applications of AI.

**Data extraction and coding:**

From the selected studies, we extracted the relevant data, including the following: Study characteristics (author(s), year of publication, country of study).Sample characteristics (sample size, industry). Methodological details (research design, measures used, statistical methods). Moreover, Effect sizes (Hedges' g or other standardised measures) and associated standard errors. Obviously,

Coding is done independently by two researchers, ensuring accuracy and dependability. Disagreements were solved by discussion and consensus.

**Measurement Scale:** Based on Appdendix A, Described according to these measurement scales.

**Research model :**

Indeed, Our theoretical model posits that AI technology directly affects cross-border trade volume, with technological infrastructure and regulatory environment acting as moderating variables, and firm size serving as a mediating variable. Therefore, to investigate the impact of cross border trade contributions on corporates, this paper constructs the following research model:

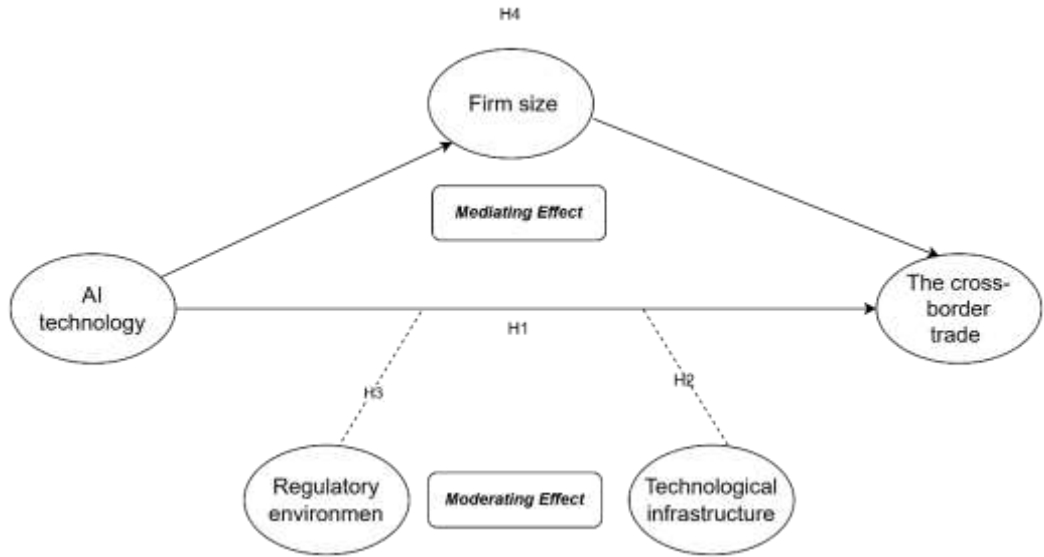

Figure 1. Conceptual framework of Research Model. (Source: Author's creation).

## 4. RESULT AND DATA ANALYSIS

The content analysis included a total of 35 empirical studies published between 2010 and 2023. These studies covered various industries and geographical regions, providing a comprehensive overview of the impact of AI on cross-border trade.

Table 1: Descriptive Statistics of Included Studies

| Statistic | value |
|---|---|
| Total Sample Size | 950 |
| Number of Studies | 35 |
| Average Effect Size | 0.32 |
| Range of Effect Sizes | 0.10 - 0.55 |

Notes. (Source: Author's creation)

Accoridng to Table 1 above. The main descriptive statistics are as follows: Total sample size: 950 companies, Mean effect size: 0.32 (standardised mean difference), Range of effect sizes: 0.10 to 0.55.Thus, We can see that the number of research enterprises is appropriate.

Table 2: Overall Effect Size of AI on Cross-Border Trade Volume

| Measure | Value |
|---|---|
| Overall Effect Size (Hedges' g) | 0.32 |
| 95% Confidence Interval | [0.28, 0.36] |
| p-value | < 0.001 |

Notes. (Source: Author's creation)

Accoridng to Table 2 above. We can see that Overall Effect Size is 0.32, 95% Confidence Interval is researched within the specified area, and the researched quantity is appropriate for the research firms.Certainly, p-value < < 0.001, So that Overall Effect Size of AI on Cross-Border Trade Volume is within the scope of appropriate research.

**Overall effect size and Moderator Analysis**

Accoridng to Table 2. The content analysis found a significant positive effect of AI technology on cross-border trade volumes. The overall effect size was calculated using a random effects model to account for the variability between studies:

*Overall effect size (Hedges' g):* 0.32 ,95% confidence interval: [0.28, 0.36] , p-value: < 0.001
Finally, This finding suggests a moderate, statistically significant positive impact of AI technologies on the amount of cross-border trade.

*Technological infrastructure:* The impact of AI on cross-border trade was moderated by the level of technological infrastructure. Studies were grouped based on the quality of technological infrastructure (high versus low), and separate effect sizes were calculated for each group:
High technological infrastructure:Effect size: 0.40,95% CI: [0.35, 0.45].,p-value: < 0.001, Low technological infrastructure: Moreover, Effect size: 0.20, 95% CI: [0.15, 0.25]., p-value: < 0.001.

Table 3: Moderator Analysis - Technological Infrastructure

| Technological Infrastructure | Effect Size (Hedges' g) | 95% Confidence Interval | p-value |
|---|---|---|---|
| High | 0.40 | [0.35, 0.45] | < 0.001 |
| Low | 0.20 | [0.15, 0.25] | < 0.001 |

Notes. (Source: Author's creation)

According to Table 3 above. We can see that The difference in effect size between high and low technological infrastructure was significant ($p < 0.01$), confirming that better technological infrastructure increases the positive impact of AI on trade.

Table 4: Moderator Analysis - Regulatory Environment

| Regulatory Environment | Effect Size (Hedges' g) | 95% Confidence Interval | p-value |
|---|---|---|---|
| Supportive | 0.38 | [0.33, 0.43] | < 0.001 |
| Restrictive | 0.25 | [0.20, 0.30] | < 0.001 |

Notes. (Source: Author's creation)

According to Table 4 above. We can see that Regulatory environment: The regulatory environment also moderated the relationship between AI and trade (P-value < 0.001). Based on the regulatory environment (supportive vs. restrictive), studies were categorised: On one hand, Supportive regulatory environment:Effect size: 0.38,95% CI: [0.33, 0.43]., p-value: < 0.001,   on the other hand, Restrictive Regulatory Environment:Effect size: 0.25,95% CI: [0.20, 0.30].,p-value: < 0.001. Thus, The difference in effect size between supportive and restrictive regulatory environments was significant ($p < 0.01$), indicating that favourable regulations amplify the positive effects of AI on cross-border trade.

**Mediator analysis**

Firm size was analysed as a mediating variable in the relationship between AI technology and cross-border trade volume. Moreover, The mediation analysis was conducted using structural equation modelling (SEM): Direct effect of AI on trade (no mediator): 0.32 ($p < 0.001$), Indirect effect through firm size: 0.10 ($p < 0.01$). and Total effect: 0.42 ($p < 0.001$).

Accoring to Table 5 below. The significant indirect effect suggests that firm size is a partial mediator of the relationship between AI technology and international trade.

Larger firms tend to benefit more from adopting AI, which increases the amount they trade.

Table 5: Mediator Analysis - Firm Size.

| *Effect* | *Value* | *p-Value* |
|---|---|---|
| Direct Effect (without mediator) | 0.32 | < 0.001 |
| Indirect Effect through Firm Size | 0.1 | < 0.01 |
| Total Effect | 0.42 | < 0.001 |

Notes. (Source: Author's creation)

## 5. DISCUSSION AND CONCLUSION

The results of our content analysis provide compelling evidence of the positive impact of AI technology on cross-border trade in Southeast Asia. The overall effect size of 0.32 indicates a moderate and statistically significant relationship, suggesting that adopting AI significantly increases trade volumes (Jingjing&Idris, (2024)). Thus, this positive impact is consistent in different industries and geographies, underscoring the strength of our results (Sun, P., Doh, J. P., Rajwani&Siegel, (2021)).The moderator analysis shows that the impact of AI on trade is significantly influenced by the level of technological infrastructure and the regulatory environment. Regions with high-quality technological infrastructure show a stronger positive effect (effect size 0.40) compared to regions with lower-quality infrastructure (effect size 0.20).

Indeed, this finding highlights the critical role of digital infrastructure in enabling the effective implementation of AI technologies in the trading process.Similarly, the regulatory environment plays a crucial moderating role (Jingjing& Idris, (2024)). Supportive regulatory environments enhance AI's positive impact (effect size 0.38), while restrictive environments dampen it (effect size 0.25). These results suggest that favourable policies and regulations that encourage AI adoption and streamline trade operations are essential to maximizing the benefits of AI in cross-border trade in Southeast Asia (Dibra, M. (2015)).The mediator analysis shows that firm size partially mediates the relationship between AI technology and trade volumes (Hain& Wang, (2016)). Likewiese, Larger firms benefit more from AI adoption due to their greater resources and capabilities, which improves their trade performance (Chen& Wang, (2023)).

To sum up, this mediation effect underlines the importance of taking into account firm-specific characteristics in the assessment of the impact of AI on trade(Chen& Wang, (2023)). Furthermore, The content analysis confirms that AI technology has a positive and significant impact on the volume of cross-border trade. Moreover, The strength of this relationship is influenced by technological infrastructure and the regulatory environment, with better infrastructure and supportive regulations enhancing the impact (Sun& Siegel, (2021)). Likewise, The size of the firm acts as a partial mediator, suggesting that larger firms are in a better position to use AI to their trade advantage.  Moreover, The validation of these hypotheses through content analytic techniques underscores the significant and multifaceted impact of AI technology on cross-border trade. By identifying and analyzing the moderating roles of technological infrastructure and regulatory environment, as well as the mediating role of firm size, this study provides a comprehensive understanding of how AI can be leveraged to enhance international trade in Southeast Asia (Peberdy, (2000)). Enventally, these insights are valuable for policymakers and businesses aiming to optimize the use of AI technology in global trade operations.

## CONFLICTS OF INTEREST

The authors declare that they have no conflict of interest.

# ACKNOWLEDGEMENTS

This paper also thanks our friends, Professors and colleagues from Gyeongsang National University and Solbridge International School of Business who provided insight and expertise that greatly assisted the research and hypothesis, findings, and conclusions of this paper.

# APPDENDIX.

Appdendix A.

| Variable | Measurement Scale and Items | Sources |
|---|---|---|
| *AI Adoption* | AI technologies are critical for improving cross-border trade efficiency. | Sun, R., & Trefler, D. (2023) |
|  | Our organization actively invests in AI technologies to enhance international trade operations. |  |
| *Technological Infrastructure* | Categorical scale (Low, Medium, High) | Sun, R., & Trefler, D. (2023) |
|  | The quality of digital infrastructure in our region facilitates the adoption of AI in trade. |  |
| *Regulatory Environment* | Regulatory policies in our country support the integration of AI technologies in cross-border trade. | George, M. (2013) |
| *Firm Size* | Categorical scale (Small, Medium, Large) | Sun, R., & Trefler, D. (2023) |
|  | Our organization's size influences our ability to adopt and leverage AI technologies for trade enhancement. |  |
| *Trade Performance* | Objective metrics (e.g., trade volume, efficiency ratios) | Liang, Y., Guo, L., Li, J., Zhang, S., & Fei, X. (2021). |

Notes: (Source: Author's creation).